\documentclass{article}

\usepackage{arxiv, xcolor}

\usepackage[utf8]{inputenc} % allow utf-8 input
\usepackage[T1]{fontenc}    % use 8-bit T1 fonts
\usepackage{url}            % simple URL typesetting
\usepackage{booktabs}       % professional-quality tables
\usepackage{amsfonts}       % blackboard math symbols
\usepackage{nicefrac}       % compact symbols for 1/2, etc.
\usepackage{microtype}      % microtypography
\usepackage{graphicx}
\usepackage{hyperref} % For clickable references
\usepackage{natbib}
\usepackage{doi}
\usepackage{float}
\usepackage{comment}
\newcommand\numberthis{\addtocounter{equation}{1}\tag{\theequation}}
\usepackage{graphicx}  % For including images
\usepackage{caption}   % For customizing captions
\usepackage{subcaption} 
\usepackage{amsmath}  % Core package for mathematical equations and environments
\usepackage{amssymb}  % Additional symbols (like \mathbb for real numbers)
\usepackage{amsfonts} % Additional fonts for math symbols
\usepackage{mathtools} % Enhancements to amsmath
\usepackage{bm} % Bold math symbols

\title{A New Perspective to Fish Trajectory Imputation: \\
A Methodology for Spatiotemporal Modeling of Acoustically Tagged Fish Data }

\author{ \href{https://orcid.org/0009-0005-5048-6795}{\includegraphics[scale=0.06]{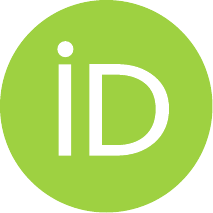}\hspace{1mm}Mahshid Ahmadian}\thanks{Corresponding author. \texttt{ahmadianm@vcu.edu}} \\
	Department of Statistical Sciences and Operations Research\\
	Virginia Commonwealth University\\
	Richmond, VA 23220 \\
	\And
	Edward L. Boone \\
	Department of Statistical Sciences and Operations Research\\
	Virginia Commonwealth University\\
	Richmond, VA 23220 \\
	\And
	Grace S. Chiu \\
	William \& Mary's Virginia Institute of Marine Science\\
	Gloucester Point, VA 23062 \\
        and \\
        Department of Statistical Sciences and Operations Research\\
	Virginia Commonwealth University\\
	Richmond, VA 23220 
}

\date{}

\renewcommand{\headeright}{Preprint}
\renewcommand{\shorttitle}{Spatiotemporal Modeling of Acoustically Tagged Fish Data}

\hypersetup{
pdftitle={Spatiotemporal Modeling of Acoustically Tagged Fish Data},
pdfsubject={Statistics},
pdfauthor={Mahshid ~Ahmadian, Edward L. ~Boone, Grace S. ~Chiu},
pdfkeywords={Animal movement, Missing not at random, Spatiotemporal data},
}

\begin{document}
\maketitle
%%%%%%%%%%%%%%%%%%%%%%%%%%%%%%%%%%%%%%%%%%%%%%%%%%%%%%%%%%%%%%%%%%%%%%%%%%%%%%%%%%%%%%%%%%%%%%%%%%%%%%%%%%%%%%%%%%%
\begin{abstract}
The focus of this paper is a key component of a methodology for understanding, interpolating, and predicting fish movement patterns based on spatiotemporal data recorded by spatially static acoustic receivers. Unlike GPS trackers which emit satellite signals from the animal's location, acoustic receivers are akin to stationary motion sensors that record movements within their detection range. Thus, for periods of time, fish may be far from the receivers, resulting in the absence of observations. The lack of information on the fish's location for extended time periods poses challenges to the understanding of fish movement patterns, and hence, the identification of proper statistical inference frameworks for modeling the trajectories.
As the initial step in our methodology, in this paper, we devise and implement a simulation-based imputation strategy that relies on both Markov chain and random-walk principles to enhance our dataset over time. This methodology will be generalizable and applicable to all fish species with similar migration patterns or data with similar structures due to the use of static acoustic receivers.
\end{abstract}
% keywords
\keywords{Animal movement, Spatiotemporal data, Missing not at random}
%%%%%%%%%%%%%%%%%%%%%%%%%%%%%%%%%%%%%%%%%%%%%%%%%%%%%%%%%%%%%%%%%%%%%%%%%%%%%%%%%%%%%%%%%%%%%%%%%%%%%%%%%%%%%%%%%%%
\section{Introduction}
Significant emphasis has been put on the study and modeling of spatiotemporal data over the years. In particular, modeling animal movement over time across a spatial landscape has gained considerable attention. For example, investigating the displacements of marine animals in water is crucial for advancing marine science, particularly understanding the population dynamics of various fish species and the effects on environmental and ecosystem conditions due to fishing practices. The structure of datasets in this type of research highly depends on the data collection method. Indeed, a statistical methodology that can generate useful predictions has been lacking for complex data structures like ours. This poses challenges that are not readily addressed by common methods, as we explain in section \ref{sec-data}.
\par
Previous research has employed diverse approaches for modeling animal movement data, such as agent-based models \cite{ref1, ref2,ref3, ref21}, Brownian bridge movement models \cite{ref25, ref26}, Lévy flight models \cite{ref4,ref5}, state-space models \cite{ref7, ref16, ref17, ref30}, Markov models \cite{ref20, ref19}, hidden Markov models \cite{ref8, ref18}, and generalized additive models \cite{ref10}. In addition, there is a wealth of research dedicated to statistical methods specifically for studying animal movement using data derived from telemetry tools where the receivers are arranged in regular spatial and/or temporal intervals so that detections are not sparse both in space and in time \cite{ref27, ref28}. 
%\\
Each approach has its own benefits and limitations. For example, agent-based models (ABMs) have been widely used, especially by ecologists and environmental scientists, to simulate the actions and interactions of individual agents (fish, in our case), and assess their effects on the population.
Brownian bridge models involve continuous-time random walks between observed locations. 
Lévy flight models extend the random walks to incorporate a heavy-tailed distribution for each defined step.
State-space models use latent and observed states probabilistically. They are often used in time series forecasting, economic indicators, and tracking and navigation. These models have been used in the study of animal movement because they are very effective in handling incomplete data, where animals are partially observed through receivers or satellites.
Moreover, hidden Markov models extend the Markov model by incorporating hidden states that represent unobserved steps in an animal movement process. Generalized additive models are used to investigate complex animal behaviors and movement patterns by offering a flexible approach to modeling non-linear relationships between animal movement and environmental factors. However, it is unclear if any
of these approaches can fully address the unique challenges posed by our acoustic telemetry data, such as having sparsely located, spatially static receivers along with irregular detection patterns over time. As such, the location of the animal is unavailable much more often than not. 
The occasional detections are dependent on receiver placement and environmental factors, and thus they require models that take into account such complexities. This necessitates the development of a new spatiotemporal model. 
\par
The purpose of this paper is to propose a new method for imputing fish trajectories, particularly when the data are collected using acoustic receivers. Many existing methods for imputing unobserved locations in an animal's trajectory rely on observed locations in the form of GPS signals emitted at regular time intervals. However, in our case, a new methodology is needed to build on the Markov property of random walks to incorporate possible behavioral characteristics due to the substantial sparsity and irregularity of observations that occur only when the animal happens to be inside the detection range of a sensor that has been pre-installed at fixed geographical locations. For our proposed approach, two key components in the movement model are the updated direction of fish movement at each step and the remaining distance until the next observed location. 
\par
The organization of this paper is as follows: Section \ref{sec-data} introduces the dataset that we used for this study and it highlights the challenges posed by the dataset structure. Section \ref{sec-model} delves into the model on which our imputation algorithm is based, followed by section \ref{sec-likelihood} which discusses the joint probability distribution associated with our method. In section \ref{sec-sim}, we present a simulation study and the visual results of the proposed algorithm. Finally, section \ref{sec-discuss} outlines the next steps and potential directions for future research.
%%%%%%%%%%%%%%%%%%%%%%%%%%%%%%%%%%%%%%%%%%%%%%%%%%%%%%%%%%%%%%%%%%%%%%%%%%%%%%%%%%%%%%%%%%%%%%%%%%%%%%%%%%%%%%%%%%%
\section{Data \label{sec-data}}
Methods for analyzing environmental data may vary across datasets, influenced not only by the nature of the scientific problem but also by the methods of data collection. Our research is tailored to a specific dataset by \cite{ref10} that provides information on fish movement along the US East Coast.
The data, collected from 2014 to 2020, utilize acoustic tagging of a species known as cobia (\textit{Rachycentron canadum}) in the Atlantic waters and include details such as the GPS coordinates of the spatially static acoustic receivers and the corresponding time stamp of the detection of a tagged fish.
Other studies are dedicated to analyzing cobia, and they have employed various techniques to understand their movement patterns and habitat utilization. Pop-up satellite tags have been used to assess the post-release survival, net travel distance, and habitat utilization of cobia in Virginia waters, providing valuable data on their movements and behavior \cite{ref22}.
However, compared to pop-up satellite archival tags, acoustic tag receivers can offer more precise and continuous tracking of fish movements, making them a better choice for detailed, long-term studies of small fish such as cobia. Hence, the study of fish movement has been completely transformed by acoustic tagging technology. These tags emit a sound signal that is recognizable by the underwater receivers, helping ecologists track tagged marine animals near the receivers.
In many scenarios where the traditional tracking methods, such as satellites, are not effective, the use of acoustic receivers is valuable \cite{ref12}. However, this technology has its own challenges, which include high costs, behavioral impacts on animals, and data gaps and missingness caused by reasons such as environmental noise or tag failure \cite{ref13, ref14, ref15}, in addition to the sparsity (often due to high costs) and static nature of receivers being used to monitor highly mobile animals.
Moreover, studies have shown that the detection power of these receivers may decrease as the tagged fish moves farther away from the receiver, even when it remains within the receiver's range of detection \cite{ref23, ref24, ref29}.

In the case of our cobia dataset, the marine science team caught the fish and implanted each with an acoustic tag. The fish was then released into the water, in which acoustic telemetry receivers with a circular detection radius of 500 meters were strategically placed by different research groups at locations that were previously identified by local fishermen as having a high likelihood of fish presence. Therefore, they are not geographically located in a regular grid pattern. Moreover, the dataset includes observations recorded at any time of the day, leading to a varying number of daily observations from none, one, or multiple detections for any given fish. 

In this paper, we focus on cobia movements around all $48$ acoustic receivers positioned in the Chesapeake Bay, Virginia with non-overlapping detection ranges. By restricting our attention to this geographical region, we can target our model at fish movements over shorter distances between detections. In addition, we concentrate on daily time steps, meaning we consider the last observed location of the fish as its assigned location for that day. This approach is chosen because the sparsity of data is lower (thus, more manageable) at this temporal scale, based on which we propose our model.
%%%%%%%%%%%%%%%%%%%%%%%%%%%%%%%%%%%%%%%%%%%%%%%%%%%%%%%%%%%%%%%%%%%%%%%%%%%%%%%%%%%%%%%%%%%%%%%%%%%%%%%%%%%%%%%%%%%
\section{Probability Model \label{sec-model}}
It is important to emphasize that the only spatial information available on each fish is whether or not it was in a receiver’s detection range, and hence, instead of the precise location of the individual fish, only the GPS coordinates of the receivers around which the fish was detected is available.
Furthermore, this sensing regime implies that no information is available when the fish is swimming outside the detection area of the network of embedded receivers.
The complexity of the acoustic tagging data, as a result, creates a two-fold problem: the imprecise location of any detected fish and a high abundance of gaps in the location of the fish. We devise and implement an imputation strategy that incorporates Markov random-walk principles with possible behavioral characteristics (e.g., determination, curiosity) to enhance the comprehensiveness of our dataset over time.
\par
Consider the movement trajectory of one fish, and let ${X}_{jt}$ be the bivariate vector of longitude and latitude of the location of fish $j$ at time step $t$, where $j=1,\cdots, J$, and $t=1, 2, 3, \cdots $. Further, let
\begin{equation}\label{eq1}
X_{jt}=
\begin{cases}
M_{jt} & \text{fish $j$ location at $t$} \in \tau _{M_j}\\
N_{jt} & \text{fish $j$ location at $t$} \in \tau _{N_j}
\end{cases}
\end{equation}
where $\tau _{N_j}$ is the set of time steps in which fish $j$ is observed by a receiver, and $\tau _{M_j}$ is the set of time points when the fish is out of detection range ($\tau_{N_j} \cup \tau_{M_j}$ covers all time points). Thus, $M_{jt}$ is the coordinates of fish $j$ beyond the detection radius $r$ (taken to be $500m$) of any receiver, and $N_{jt}$ is the coordinates of fish $j$ within the detection radius $r$ of some receiver. Thus, letting $R_i$ denote the coordinates of receiver $i$, we have the constraints
\[
\| N_{jt} - R_i \| \leq r \quad \text{for some } i \quad \text{and} \quad \| M_{jt} - R_i \| > r \quad \text{for all } i = 1, \cdots, I
\]
\par
Imputing the location of the fish for times $t \in \tau_{N_j}$ is to pick a random point from a circular cloud of points centered at the receiver's coordinates, where the density of points decreases as the fish moves away from the center. This approach reflects the decreasing detection power of receivers discussed in section \ref{sec-data}. Hence, we assume the bivariate normal distribution

\begin{equation}\label{eq2}
    N_{jt} \sim \mathbf{N}_{2} (R_i , \Sigma_{R}), \quad \text{for some } i
\end{equation}
in which the covariance matrix is defined as:
\begin{equation}\label{eq3}
\Sigma_{R} = 
\begin{bmatrix}
\sigma _{R} ^{2} & 0\\
0 & \sigma _{R} ^{2} 
\end{bmatrix}.
\end{equation}
This means that the imputed coordinates of the fish within the detection radius of a receiver are normally distributed with an equal variance of $\sigma_{R}^{2}$ in any direction around the receiver. 
\par
However, for the time steps of $t \in \tau_{M_j}$, we use a more complex model that incorporates direction and distance. This approach is tailored to the movement characteristics of each fish and ensures that the fish moves toward the final location in a logical time frame, as follows. 
\par
Consider fish $j$ during an unobserved period of time between observations at $t= 1$ and $t= T_{j}$. That is, receivers detected fish $j$ at times $1$ and $T_j$, but not at $t=2, 3, \cdots, T_{j-1}$. By letting $\theta_{jt}$ be the angle between the fish's imputed location at time $t-1$ and its final location, say $X_{j T_{j}} = X^{*}_{j}$ imputed by equations \ref{eq2} -- \ref{eq3}, the location at each step $t$ for $t = 2, 3, \cdots, T_{j}-1$ is modeled as
\begin{equation}\label{eq4}
    M_{jt} = X_{j(t-1)} +
    \begin{bmatrix}
|D_{jt}|  \cos( \theta_{jt} + \psi_{jt} ) \\
|D_{jt}|  \sin( \theta_{jt} + \psi_{jt} )
\end{bmatrix},
\end{equation}
where
%$X_{j(t-1)}$ is the imputed location of fish at time $t-1$ that was imputed by either of the model levels.
\begin{gather}
D_{jt} \sim \mathcal{N} (\frac{d_{jt}}{T_{j} - (t-1)} ,  {d_{jt}^{\phi}}),  \quad
\psi_{jt} \sim \mathcal{N} (0 , \sigma_{\psi_{jt}} ^{2}), \\
d_{jt} = Max \{ (\rVert X_{j}^{*} - X_{j(t-1)} \rVert - 2r), (\rVert X_{j}^{*} - X_{j(t-1)} \rVert) \}, \\
    \sigma_{\psi_{jt}} ^{2}=
    \begin{cases}
        \gamma . \exp(\alpha (n_{j}-(t-1)) )& n_{j}\leq \beta\\ 
        \gamma . U_t & n_{j}> \beta
    \end{cases}, \label{eq5}
\end{gather}
and the quantities in equations \ref{eq4} -- \ref{eq5} are described as follows. 
%The total number of unobserved time steps is defines as $n_j=T_j-2$}

We let $D_{jt}$ be a latent stochastic quantity that determines the distance between $X_{j(t-1)}$ and $M_{jt}$. To accurately account for the distance traveled at each time step, we model $D_{jt}$ as normally distributed where the mean is defined to ensure the imputation process considers the remaining distance. 
Here, $d_{jt}$ represents the total distance that remains to be traveled from time $t$ minus the two receivers' radii.
Additionally, the variance of $D_{jt}$ is modeled using a power decay function to capture variability in the remaining distance traveled, where $\phi$ is a parameter that controls the variance's decay rate with respect to the remaining distance. (This is a decay function because $d_{jt}$ decreases as $t$ increases.)
%\\
This formulation ensures that both the mean and the variance of $D_{jt}$ dynamically adjust based on the remaining distance, providing a robust framework for imputing missing location in a spatiotemporal context.
\par
For the angle at which the fish travels from $X_{j(t-1)}$ to $M_{jt}$, we incorporate a direction of movement and reflect the fish's determination toward $X_j^\ast$ (the end point of its trajectory) through $\gamma$ in equation \ref{eq5}; it represents the initial value of the variance of the noise term $\psi_{jt}$ associated with $\theta_{jt}$, and the parameter $\alpha$ that controls the exponential decay of the variance. The threshold parameter $\beta$ determines the point at which the variance model switches from exponential to a uniform distribution, where $U_t \sim$ Uniform$(0,1)$.
This assumption implies that for smaller gaps ($n_{j}\leq \beta$, where $n_j=T_j-2$ is the total number of unobserved time steps) the variance decreases exponentially as the fish moves toward $X_j^\ast$. However, for larger time gaps ($n_{j}> \beta$) we give extra freedom to $\psi_{jt}$ by considering a variance value derived from the uniform distribution to reflect the fish's curiosity that results in the longer time between detections.
%%%%%%%%%%%%%%%%%%%%%%%%%%%%%%%%%%%%%%%%%%%%%%%%%%%%%%%%%%%%%%%%%%%%%%%%%%%%%%%%%%%%%%%%%%%%%%%%%%%%%%%%%%%%%%%%%%%%%%%%%%%%%%%%%%%%%%%%%%%%
\section{Likelihood \label{sec-likelihood}}
For any given fish from the dataset, the above probability model can be used to simulate its daily locations between two consecutive detections, forming a single imputed segment of its full trajectory. Connecting all imputed segments gives rise to the full imputed trajectory for the fish. The stochastic nature of the imputed trajectory means that one simulated trajectory may be more probable than another, even if both trajectories are simulated for the same fish from the same model. In this section, we derive the likelihood for any trajectory that is simulated from the model in section \ref{sec-model}. The likelihood will be used to display the $90 \%$ most likely paths in the form of a heatmap in section \ref{sec-sim}, thus allowing us to visualize the geographical areas over which the fish's presence is most likely to be concentrated.

To derive the likelihood, let us consider fish $j$ and one segment of this fish trajectory which started at $t=1$ and ended at $t=T_j$. Thus, we drop the subscript $j$ from the following equations.
\par
According to equation \ref{eq1}, $\forall \quad t \in \tau_{N} = \{1, T \}$, 
\begin{equation*}
X_{t} =
\begin{bmatrix}
X_t^{(1)} \\
X_t^{(2)}
\end{bmatrix}
\sim \mathbf{N}_2({R}_t, \Sigma_R)
\end{equation*}
where 
\begin{equation*}
    {R}_t=
    \begin{cases}
       R_i & \textit{if receiver i detected the fish at time t}\\ 
        $NA$ & \textit{otherwise}
    \end{cases}.
\end{equation*}
Therefore, the joint density of $X_1$ and $ X_T$ is defined as

\begin{equation*}
 \mathcal{P}_0 ({X}_1,{X}_T | {R}_1, {R}_T, \sigma^2_R) = \left( \frac{1}{2\pi \sigma_R^{2}} \right)^2
  \exp\left( -\frac{1}{2\sigma_{R}^2} \sum_{t=1,T}\left[
(X_t^{(1)} - R_{t}^{(1)})^2 + (X_t^{(2)} - R_{t}^{(2)})^2 \right] \right).
\end{equation*}

Let $t$ belong to the set  $ \tau_{M} = \{2, 3, \cdots, T-1\}$, and $\textbf{Z}_{t} =
\begin{bmatrix}
\cos( \theta^*_{t} ) \\
\sin( \theta^*_{t} )
\end{bmatrix}$,
for $\theta^*_{t} = \theta_{t} + \psi_{t}$; hence,
\begin{equation*}
\theta^*_{t} | \theta_{t}, \sigma_{\psi_{t}}^2 \sim \mathcal{N}(\theta_{t}, \sigma^{2}_{\psi_{t}}).
\end{equation*}
By using a two-term Taylor expansion of $\textbf{Z}_t = \textbf{Z}_t(\theta^*_{t} )$ around $\theta_{t}$, we have (see Appendix \ref{Ap1})
$$\textbf{Z}_{t} | \theta_{t}, \sigma_{\psi_{t}}^2 \overset {approx} \sim  \mathbf{N}_{2} (\boldsymbol{\mu}_{z}, \Sigma_{z})$$
where
\begin{equation*}
\boldsymbol{\mu} _{z} =
\begin{bmatrix}
\cos( \theta_{t} ) - \frac{1}{2} \cos( \theta_{t}) \sigma^{2}_{\psi_{t}}  \\
\sin( \theta_{t}) - \frac{1}{2} \sin( \theta_{t}) \sigma^{2}_{\psi_{t}}
\end{bmatrix},
\end{equation*}
and letting $\mathrm{C}_t = \cos( \theta_{t})$ and $\mathrm{S}_t = \sin( \theta_{t})$, the covariance matrix is
\begin{equation*}
\Sigma_{z} =  \sigma^{2}_{\psi_{t}}
\begin{bmatrix}
\mathrm{S}_t ^{2} + \frac{1}{2} \mathrm{C}_t ^{2} \sigma^{2}_{\psi_{t}} &
-\mathrm{S}_t\mathrm{C}_t + \frac{1}{2}\mathrm{S}_t\mathrm{C}_t \sigma^{2}_{\psi_{t}}\\
-\mathrm{S}_t\mathrm{C}_t + \frac{1}{2}\mathrm{S}_t\mathrm{C}_t \sigma^{2}_{\psi_{t}} &
\mathrm{C}_t ^{2}  + \frac{1}{2} \mathrm{S}_t ^{2} \sigma^{2}_{\psi_{t}}
\end{bmatrix}.
\end{equation*}
Next, let $\textbf{W}_t = |D_t|\textbf{ Z}_t $, thus 
\begin{equation*}
\textbf{W}_{t} | D_t, \theta_t, \sigma^{2}_{\psi_{t}} \overset {approx} \sim \mathbf{N}_{2} \left( |D_t|\boldsymbol{\mu}_{z}, D_t^2 \Sigma_{z} \right).
\end{equation*}
Letting ${X}_t = {X}_{t-1} + W_t$, we have, 
\begin{equation*}
{X}_t | {X}_{t-1},  D_t, \theta_t, \sigma^{2}_{\psi_{t}} \overset {approx} \sim \mathbf{N}_{2} \left( {X}_{t-1} + |D_t|\boldsymbol{\mu}_{z}, D_t^2 \Sigma_{z} \right)
\end{equation*}
with density
\begin{align*}
&\mathcal{P}_{1}({X}_t|{X}_{t-1}, D_t, \theta_t, \sigma^{2}_{\psi_{t}} ) = \\
& \frac{1}{2\pi D_t^2 |\Sigma_z|^{1/2}} \exp \left( -\frac{1}{2} ({X}_t - {X}_{t-1} - |D_t| \boldsymbol{\mu}_z)^T (D_t^2 \Sigma_z)^{-1} (X_t - {X}_{t-1} - |D_t| \boldsymbol{\mu}_z ) \right).
\end{align*}
\\
Next, as $\phi$ is non-stochastic, we have $D_t | d_t , \phi \sim \mathcal{N} (\frac{d_{t}}{T - (t-1)}, {d_{t}^{\phi}}) $ with density
\begin{equation*}
\mathcal{P}_2 (D_t | d_t , \phi) =
\frac{1}{\sqrt{2 \pi d_{t}^{\phi} }} \exp\left( -\frac{1}{2 d_{t}^{\phi}} \left( D_t - \frac{d_{t}}{T - (t-1)}\right)^2   \right).
\end{equation*}
Note that conditioned on ${X}_{t-1}$ and $X^*$, both $d_t$ and $\theta_t$ are non-stochastic, and depending on $n_j$, $ \sigma^{2}_{\psi_{t}}$ is either non-stochastic or has a uniform distribution. Thus, the joint distribution of stochastic terms relevant to time $ t \in \tau_M$ is: %by assuming $\Theta = \{ \psi_t, \sigma^2_{\psi}, \phi \}$
\begin{align*}
&f({X}_t, {X}_{t-1}, {X}_1, {X}^*, D_t, d_t, \theta_t) = \\
& \mathcal{P}_{1}({X}_t| {X}_{t-1}, {D}_{t}, \sigma^{2}_{\psi_{t}}, \theta_t)\ 
\mathcal{P}_2 (D_t | d_{t} , \phi)\ 
f({X}_{t-1}, {X}^*).
%\times &\mathcal{P} (d_t , \theta_t |, \textbf{X}_1, \textbf{X}^*, D_{t-1}, d_{t-1})
\end{align*}

\par
We expand it to all $t \in \tau_N \cup \tau_M$ (see Appendix) to have
\begin{align*}
    & f(X_1, X_2, \cdots, X_{T-1}, X^*, D_2, \cdots, D_{T-1}, \theta_2, \cdots, \theta_{T-1},
    d_2, \cdots, d_{T-1}) =\\
    & \mathcal{P}_0 (X_1,X^*| R_1, R_T, \sigma^2_R)  \prod_{t=2}^{T-1} 
    \mathcal{P}_{1}(X_t|X_{t-1}, D_t, \theta_t, \sigma^{2}_{\psi_{t}})\ \mathcal{P}_2 (D_t | d_t , \phi). \numberthis \label{eq6}
\end{align*}

The function in equation \ref{eq6} above pertains to one single segment of the full trajectory of the fish.
Therefore, with all $K$ segments of the trajectory, the full joint distribution is
\begin{align}\label{eq9}
    & f(X_{k_1}, X_{k_2}, \cdots, X_{k_{T-1}}, X^{*}_k, D_{k_2}, \cdots, D_{k_{T-1}}, \theta_{k_2}, \cdots, \theta_{k_{T-1}},
    d_{k_2}, \cdots, d_{k_{T-1}}) = \\
    & \prod_{k=1}^{K} \left( \mathcal{P}_0 (X_{k_1},X_{k_T} | R_{k_1}, R_{k_t}, \sigma^2_R)
    \prod_{t=2}^{T-1}   \mathcal{P}_{1}(X_{k_t}|X_{k_{t-1}}, D_{k_t}, \theta_{k_t}, \sigma^{2}_{\psi_{t}}) \ \mathcal{P}_2 (D_{k_t} | d_{k_t}, \phi) \right)
\end{align}
where $k_t$ denotes the $t$-th time step of the $k$-th segment of the fish trajectory, and $X_{k_1} = X_{k-1}^*$ for all $k>1$, i.e. the fish's new starting location is its previous end location.
%%%%%%%%%%%%%%%%%%%%%%%%%%%%%%%%%%%%%%%%%%%%%%%%%%%%%%%%%%%%%%%%%%%%%%%%%%%%%%%%%%%%%%%
%%%%%%%%%%%%%%%%%%%%%%%%%%%%%%%%%%%%%%%%%%%%%%%%%%%%%%%%%%%%%%%%%%%%%%%%%%%%%%%%%%%%%%%
\section{Simulation Study \label{sec-sim}}
We conduct a simulation study to validate the proposed imputation model. To this end, we apply the method to two selected fish with acoustic tag numbers $18453$ and $18434$. One fish exhibited migratory behavior, while the other fish's recorded movement was almost stationary. The timeline for each fish varies. The first fish's time period is from September 1, 2018, to September 12, 2018, while the second fish's period is from September 23, 2017, to October 10, 2017. Since no environmental covariates or population interactions are considered in this current model, and these two fish trajectories are imputed independently, the difference in timelines has no effect on our imputations. The time points are currently considered the daily locations of the fish. Specifically, the first fish was detected at time points of $t = 1, 5, 10, 12$, and by receiver numbers $44, 31, 18, 13$. This indicates that the fish was detected at these four receivers over four different days within a span of twelve consecutive days, and thus, 
$K=3$, $T_{k=1} =5$, $T_{k=2} = 6$, and $T_{k=3} = 3$. 
Similarly, the second fish was detected at $t = 1, 7, 18$, and by receivers $43, 29, 25$
$K=2$, $T_{k=1} =7$, and $T_{k=2} = 12$.
\par
We demonstrate the effectiveness of the model for these different fish behaviors using parametric bootstrap sampling along with likelihood filtering to obtain the most credible estimated paths \cite{ref11}.
Moreover, to enhance realism, we simulate all non-stochastic quantities in equations \ref{eq4} to \ref{eq5} from relatively uninformative prior distributions instead of fixing each at a predetermined value. While we do not restrict the prior distributions to take on any specific parameter value, we carefully select hyperparameters that meet our desired criteria. Thus, we select prior distributions for the parameters that reflect the prior knowledge and theoretical considerations regarding the described method. For $\alpha$, we use $\Gamma(10,10)$ to ensure a positive value with a mean of 1, indicating weak prior informativeness. For $\phi$, a \textit{Lognormal}$(0.5,100)$ is used to achieve a positive value with a wide range of variability. The $\gamma$ parameter follows a \textit{Lognormal}$(2,1)$ distribution, centered at 1 to have a reasonable initial value for $\sigma^2_{\psi_{t}}$.  Lastly, $\sigma_R^2$ is modeled as \textit{inv-Gamma}$(3, 0.00298)$ to achieve a variance value that ensures $3 \mathbb{E}(\sigma_R) \approx r$. In other words, the variance accounts for the sparsity between receivers. The threshold parameter, $\beta$, is currently set at a fixed value of $3$, because according to the fish direction's variability in equation \ref{eq5} and exploratory data analysis, we believe that for a gap of length $3$ or less, the fish direction is more directly predictable compared to longer gaps. In other words, a fish has more mobility in different directions for a gap of length $4$ and more.
\par
We initially focus on the first fish (ID 18453) with relatively short time gaps.
The fish's trajectory starts at $t=1$ near receiver $44$, then $3$ days later, near receiver $31$ and then the trajectory continues to the last location at $t=12$ near receiver $13$.

\par
\begin{figure}[ht]
    \centering
    \begin{subfigure}[t]{0.45\textwidth}
        \centering
        \includegraphics[width=\textwidth]{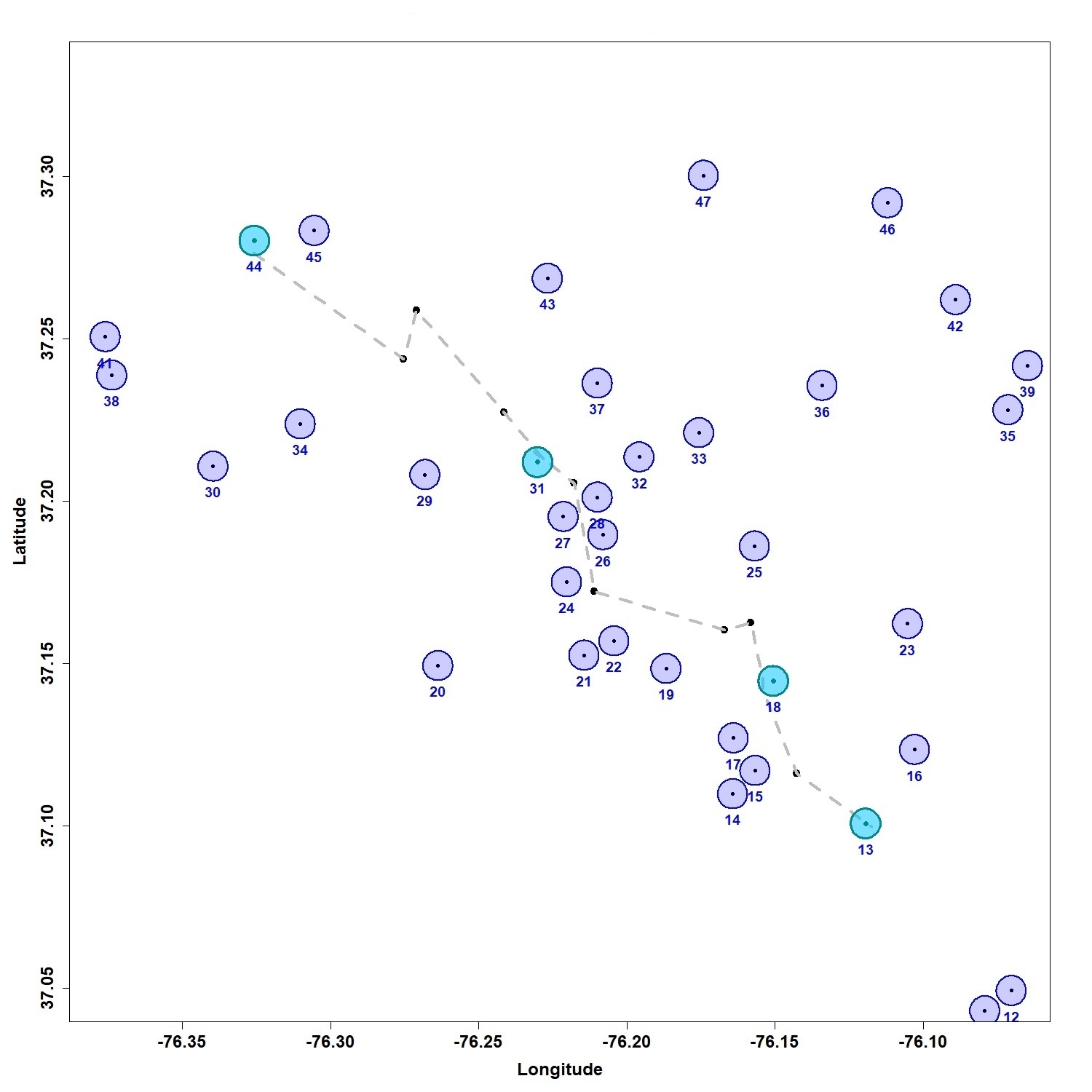}
        \caption{}
        \label{fig2a}
    \end{subfigure}
    \hfill
    \begin{subfigure}[t]{0.45\textwidth}
        \centering
        \includegraphics[width=\textwidth]{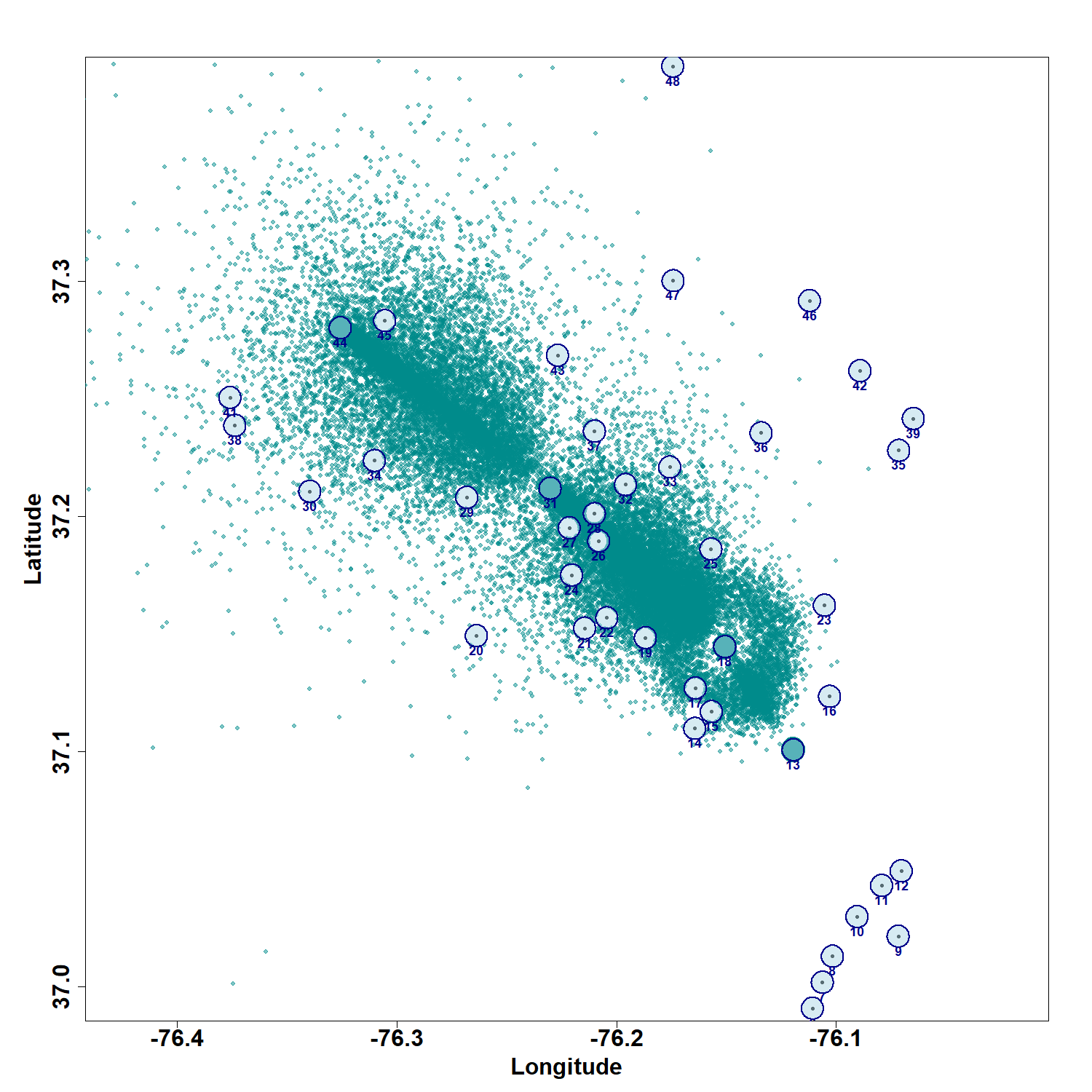}
        \caption{}
        \label{fig2b}
    \end{subfigure}
    \caption{Comparison of one simulated trajectory (panel (a)) and 4500 simulated trajectories (panel (b)) for fish ID 18453 over 12 time steps  using the proposed algorithm. The receivers that detected the fish are shown in color.}
    \label{fig:fish_trajectory_comparison}
\end{figure}

To illustrate the algorithm, we consider the first fish for one iteration (Figure \ref{fig2a}). The large black points in this figure show the fish's simulated locations at the missing time points. The gray lines connecting these points illustrate the trajectory progression from receiver $45$ to receiver $13$. Thus, it is important to note that this line does not imply a linear path between the simulated points, but is intended to show a visual guide to understand the overall movement.

\par
For example, by parametric bootstrap sampling along with likelihood filtering, we repeat the simulation algorithm for $5000$ iterations and keep $4500$ of the most probable ones (i.e. with $90 \%$ highest likelihood values). Hence, Figure \ref{fig2b} shows $4500$ iterations for fish ID $18453$. Additionally, in Figure \ref{fig3}, a heatmap of the same iterations is shown. This heatmap is available in the Appendix \ref{Ap3} with a higher resolution.

%%%%%%%%%%%%%%%%%%%%%%%%%%%%%%%%%%%%%%%%%%%%%%%%%%%%%%%%%%%%%%%%%%%%%%%%%%%%%%%%%%%%%%%%%%%%%%%%%%%%%%%%%%%%%%%%%%%

\begin{figure}[H]
    \centering
    \includegraphics[width=0.5\columnwidth]{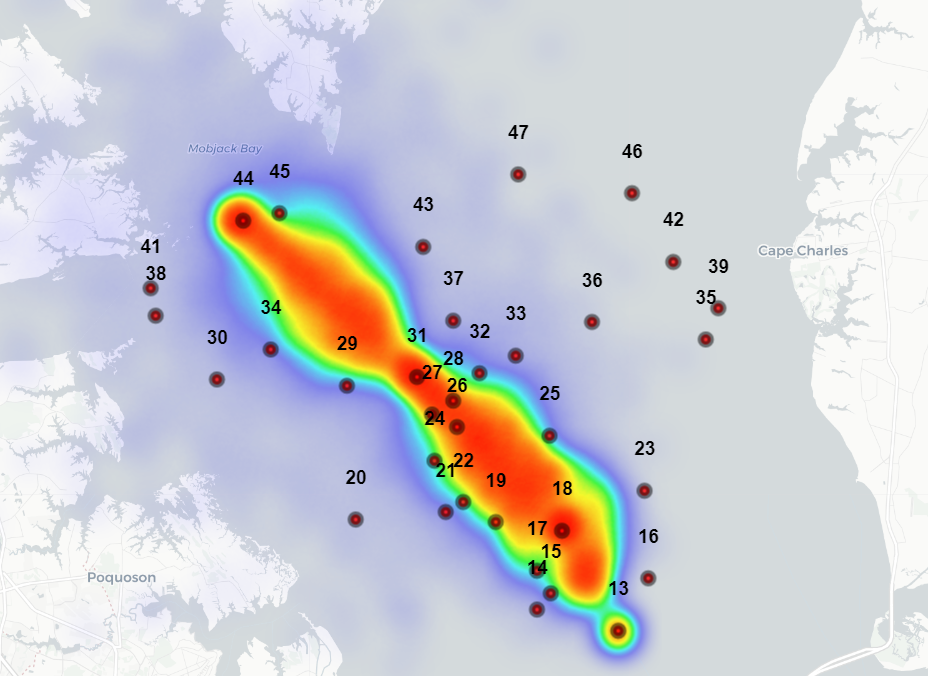}
    \caption{Heatmap corresponding to Figure 1b.}
    \label{fig3}
\end{figure}

\begin{figure}[H]
    \centering
    \begin{subfigure}[t]{0.45\textwidth}
        \centering
        \includegraphics[width=\textwidth]{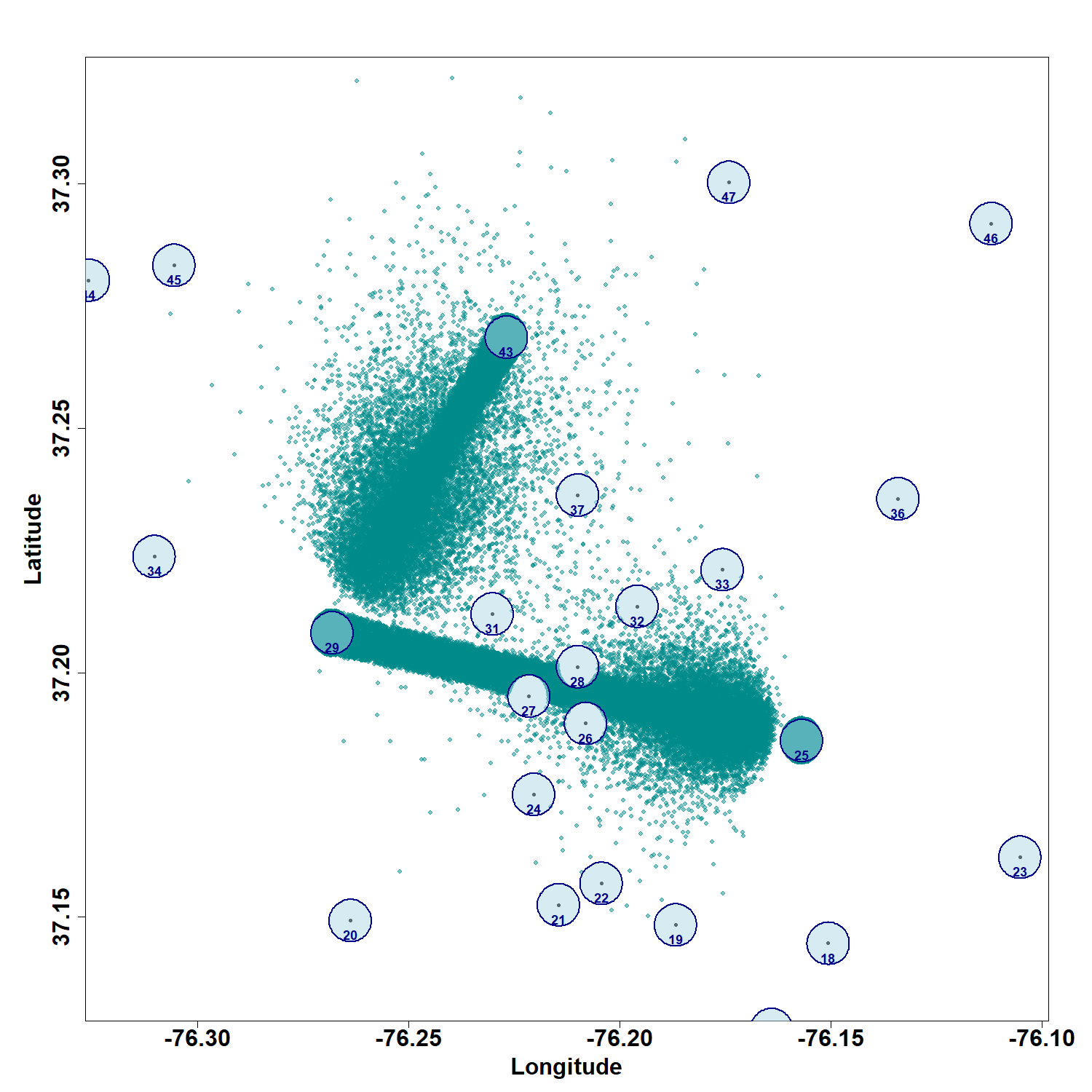}
        \caption{}
        \label{fig4a}
    \end{subfigure}
    \hfill
    \begin{subfigure}[t]{0.45\textwidth}
        \centering
        \includegraphics[width=\textwidth]{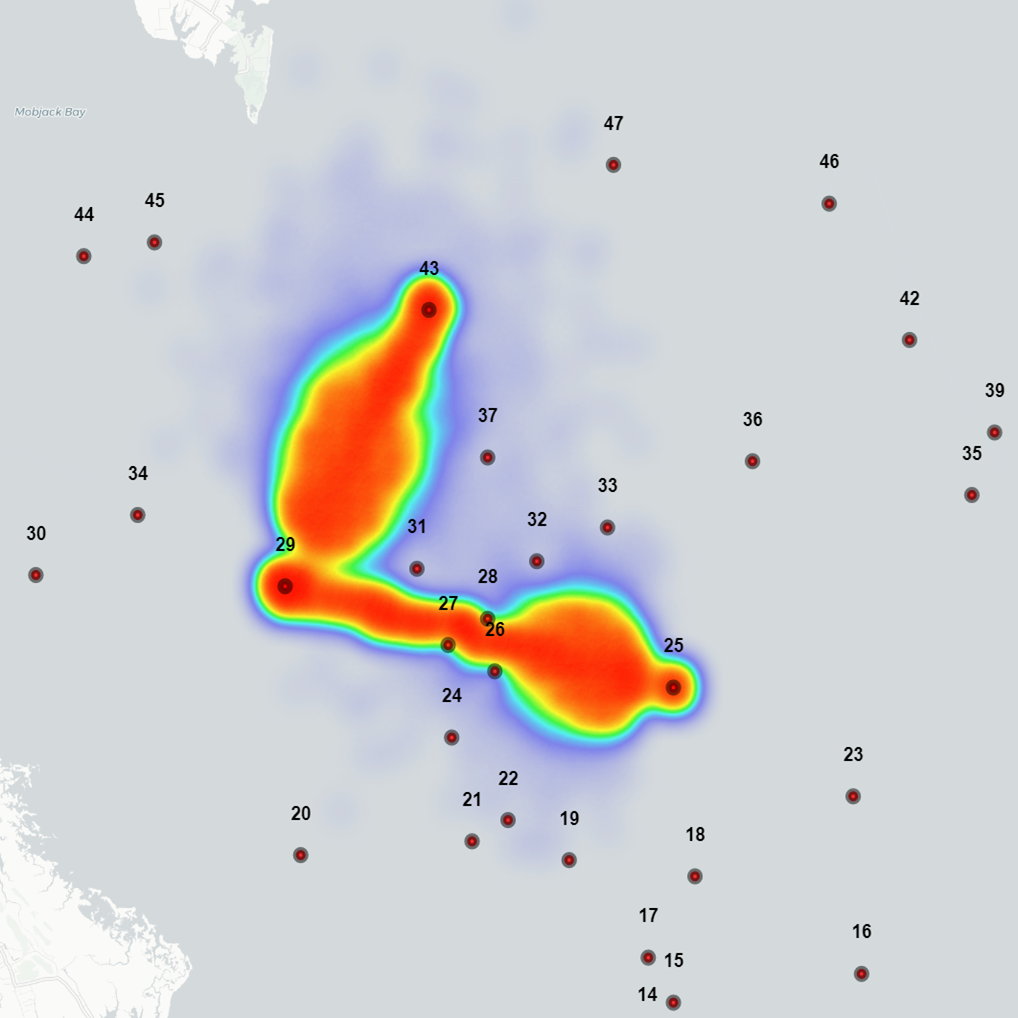}
        \caption{}
        \label{fig4b}
    \end{subfigure}
    \caption{Simulated trajectories and heatmap. Panel (a): 4500 simulated trajectories for fish ID 18434 over 18 time steps using the proposed algorithm. Panel (b): Heatmap corresponding to panel (a).}
    \label{fig:fish_simulation}
\end{figure}

The imputation for the second fish (ID $18434$) is shown in Figures \ref{fig4a} and \ref{fig4b}. Figure \ref{fig4a} shows the $90 \%$ most probable simulated trajectories out of $2000$, while Figure \ref{fig4b} shows the heatmap of the same trajectories. These two figures demonstrate the reduced realism of the proposed method when larger time gaps exist. Specifically, for the $10$ step gap between receivers $30$ and $26$, the imputed trajectories tend to follow a more linear pattern. Ecologically, cobia are known for their tendency of migratory behavior, and therefore, this simulated linearity is less realistic \cite{ref10}. During longer gaps between the recorded detections, fish are expected to exhibit non-linear, multi-directional movements rather than straight paths. Hence, this example demonstrates that our current imputation method requires modification to capture the natural movement behaviors of fish over long unobserved time periods.

\section{Discussion \label{sec-discuss}}
In summary, this paper focuses on the imputation of cobia movement trajectories while accounting for physical limitations due to the tracking technology (acoustic tagging), accompanied by visual validation.
Our imputation approach relies on simulating from bespoke probability distributions tailored to the data collection technology and animal species of interest.
The challenges addressed in this paper are the structure of the data and the methods used for data collection temporally. The data are very sparse, and the actual fish location is unknown even when presence is detected. In fact, these are the complexities of tracking the fish acoustically, which lead to a lack of applicability of the common models for animal movement analysis. In our paper, we developed a new strategy to handle these challenges.
\par
We intend to apply our algorithm to impute all fish trajectories from the full dataset ($N=67$ fish) by \cite{ref10}. The goal is to develop an inference model that also integrates relevant environmental characteristics along with the imputed data. This will improve understanding and prediction of the behavior of this fish species as well as those animals with similar movement patterns.
\par
The hierarchically structured algorithm introduced in this paper offers a framework for independently imputing short-time gaps in the trajectory of each fish. However, the individuality of fish-specific imputations and the unavailability of the fish's actual locations do not provide the context for performing a cross-validation study. 
\par
As future work, we will extend this model by incorporating a transition probability matrix for each time step, giving us the probabilities of fish transitioning between different receivers, and therefore, the fish location predictions will allow population-level inference on which k-fold cross-validation can be employed. Additionally, we will improve the fish-specific model's applicability to more extensive time gaps by incorporating multiple layers of mixture modeling to distinguish between ``lazy'' and ``nosy'' behavior, and between resident and migratory populations.
\par
Despite the progress in modeling techniques, several challenges remain. Data quality and quantity can be compromised by tag malfunctions or environmental interference, leading to incomplete datasets \cite{ref12, ref13, ref14, ref15}; these challenges are beyond the scope of our current work. Environmental factors such as water temperature and salinity influence fish movement and add complexity to model predictions. Furthermore, ensuring model accuracy and validation is essential to represent actual movement patterns accurately. Our inference framework (in progress) aims to incorporate these elements into the imputation method presented in this paper.
We anticipate our current paper and inference framework (upon completion) to add substantial rigor to this field of research.
%Additionally, existing research has examined seasonal movement patterns, and habitat use along the Texas coast, highlighting the need for better stock structure understanding to inform future fishery management (Sportfish Center, 2021). Individual-based models (IBMs) have also been developed to simulate fish movement trajectories in relation to hydraulic variables, demonstrating the applicability of such models to different species and environments (MDPI, 2018).

%%%%%%%%%%%%%%%%%%%%%%%%%%%%%%%%%%%%%%%%%%%%%%%%%%%%%%%%%%%%%
\section*{Acknowledgement}
We gratefully acknowledge the financial support of the Virginia Sea Grant through a graduate fellowship awarded to M.A. which made this research possible. We also extend our appreciation to Dr. Daniel P. Crear (Inter-American Tropical Tuna Commission) and Dr. Kevin Weng (Virginia Institute of Marine Science) for providing the necessary resources and their guidance during this work. We thank the handling editor for helpful comments that improved our paper.

\bibliographystyle{plain}
\bibliography{references}  %%% Uncomment this line and comment out the ``thebibliography'' section below to use the external .bib file (using bibtex) .

\begin{appendix}
\section{Second-order Taylor Expansion for \textbf{Z}}\label{Ap1}
For $\textbf{Z} =
\begin{bmatrix}
Z^{(1)} = \cos( \theta^* ) \\
Z^{(2)} = \sin( \theta^* )
\end{bmatrix}$,

\begin{align*}
    Z^{(1)} &= \cos \theta^* \\
    &\approx \cos \theta - (\sin \theta)(\theta^* - \theta) - \frac{1}{2} (\cos \theta)(\theta^* - \theta)^2 \\
    &\approx (\sin \theta)(\theta^* - \theta) - \cos \theta \left( 1 - \frac{(\theta^* - \theta)^2}{2} \right), \\
    \mathbb{E}(Z^{(1)}) &\approx \cos \theta \left( 1 - \frac{\sigma_\psi^2}{2} \right), \\
    \text{Var}(Z^{(1)}) &\approx (\sin \theta) ^{2} \sigma^{2}_{\psi}  + \frac{1}{2} (\cos \theta)^{2} \sigma^{4}_{\psi} \,,
\end{align*}
and
\begin{align*}
    Z^{(2)} &= \sin \theta^* \\
    &\approx \sin \theta + (\cos \theta)(\theta^* - \theta) - \frac{1}{2} (\sin \theta)(\theta^* - \theta)^2 \\
    &\approx \cos \theta (\theta^* - \theta) - (\sin \theta) \left( 1 - \frac{(\theta^* - \theta)^2}{2} \right), \\
    \mathbb{E}(Z^{(2)}) &\approx (\sin \theta) \left( 1 - \frac{\sigma_\psi^2}{2} \right), \\
    \text{Var}(Z^{(2)}) &\approx (\cos \theta) ^{2} \sigma^{2}_{\psi}  + \frac{1}{2} (\sin \theta)^{2} \sigma^{4}_{\psi} \,,
\end{align*}
and by letting $ \mathrm{C} = \cos( \theta)$ and $\mathrm{S} = \sin( \theta)$, we have
\begin{equation*}   
    \text{Cov}(Z^{(1)}, Z^{(2)}) \approx 
    -\mathrm{S} \mathrm{C} \sigma^{2}_{\psi} + \frac{1}{2}\mathrm{S} \mathrm{C} \sigma^{4}_{\psi} \,,\
\end{equation*}
\begin{align*}
    \begin{bmatrix}
        Z^{(1)} \\
        Z^{(2)}
    \end{bmatrix}
    \overset {approx} \sim \mathbf{N}_{2}
    \left(
\begin{bmatrix}
\cos( \theta ) - \frac{1}{2} \cos( \theta) \sigma^{2}_{\psi}  \\
\sin( \theta) - \frac{1}{2} \sin( \theta) \sigma^{2}_{\psi} 
\end{bmatrix}, 
\sigma^{2}_{\psi}
\begin{bmatrix}
\mathrm{S} ^{2} + \frac{1}{2} \mathrm{C}_t ^{2} \sigma^{2}_{\psi} &
-\mathrm{S} \mathrm{C}  + \frac{1}{2}\mathrm{S} \mathrm{C} \sigma^{2}_{\psi}\\
-\mathrm{S} \mathrm{C} + \frac{1}{2}\mathrm{S} \mathrm{C} \sigma^{2}_{\psi} &
\mathrm{C} ^{2}   + \frac{1}{2} \mathrm{S} ^{2} \sigma^{2}_{\psi}
\end{bmatrix}
    \right).
\end{align*}

\section{Joint Distribution of Stochastic Terms for \texorpdfstring{$T=4$}{T=4}}\label{Ap2}
Assume the trajectory lacks observations at time steps $t=2,3$. Thus, $X_4 = X^*$, and 
\begin{equation*}
\begin{aligned}
    & \mathcal{P} \left[
    \begin{matrix}
        X_1, & X_2, & X_3, & X_4, & \theta_2, & \theta_3, & D_2, & D_3, & d_2, & d_3
    \end{matrix}
    \right] \\
    = & \mathcal{P} \left[
    \begin{matrix}
        X_3 \, \vert & X_1, & X_2, & X_4, & \theta_2, & \theta_3, & D_2, & D_3, & d_2, & d_3
    \end{matrix}
    \right] \times \\
     & \mathcal{P} \left[
    \begin{matrix}
        D_3 \, \vert & X_1, & X_2, & X_4, & \theta_2, & \theta_3, & D_2, & d_2, &d_3
    \end{matrix}
    \right] \times \\
     & \mathcal{P} \left[
    \begin{matrix}
        d_3 \,, & \theta_3 \, \vert & X_1,& X_2, & X_4, & \theta_2, & D_2, & d_2
    \end{matrix}
    \right] \times \\
         & \mathcal{P} \left[
    \begin{matrix}
        X_2 \, \vert & X_1, & X_4, & \theta_2, & D_2, & d_2
    \end{matrix}
    \right] \times \\
     & \mathcal{P} \left[
    \begin{matrix}
        D_2 \, \vert & X_1, & X_4, & \theta_2, & d_2
    \end{matrix}
    \right] \times \\
     & \mathcal{P} \left[
    \begin{matrix}
        d_2 \,, & \theta_2 \, \vert & X_1, & X_4
    \end{matrix}
    \right] \times \\
     & \mathcal{P}\left[
    \begin{matrix}
        X_1
    \end{matrix}
    \right]
    \mathcal{P} \left[
    \begin{matrix}
        X_4
    \end{matrix}
    \right] \\
    = &   \mathcal{P} \left[
    \begin{matrix}
         X_3 \, \vert & X_2, & D_3, & \theta_3
    \end{matrix}
    \right]
    \mathcal{P} \left[
    \begin{matrix}
         D_3 \, \vert & d_3
    \end{matrix}
    \right] \times \\
    &\mathcal{P} \left[
    \begin{matrix}
         X_2 \, \vert & X_1, & D_2, & \theta_2
    \end{matrix}
    \right]
    \mathcal{P} \left[
    \begin{matrix}
         D_2 \, \vert & d_2
    \end{matrix}
    \right] \times \\
         & \mathcal{P}\left[
    \begin{matrix}
        X_1
    \end{matrix}
    \right]
    \mathcal{P} \left[
    \begin{matrix}
        X_4
    \end{matrix}
    \right]. \\
\end{aligned}
\end{equation*}

Note that $\mathcal{P} \left[
        d_3, \theta_3 \, \vert  X_1, X_2,  X_4,  \theta_2,  D_2,  d_2
    \right] = \mathcal{P} \left[ d_3, \theta_3 \, \vert  X_2, X_4 \right]$
drops out of the likelihood due to non-stochasticity. The same applies to $\mathcal{P} \left[ d_2, \theta_2 \, \vert  X_1, X_4 \right]$.

\section{High resolution heat map}\label{Ap3}
Download Figure \ref{fig3} in high resolution HTML format at %
%\url{https://rb.gy/z0p8h1}
\url{https://go.wm.edu/NYq32h}.

\end{appendix}

\end{document}